\def\.#1{\mathaccent 95#1}
\def\^#1{\mathaccent 94 #1}
\def\~#1{\mathaccent "7E #1}

\def\equal{\enskip =\enskip}
\def\plus{\enskip +\enskip}
\def\minus{\enskip -\enskip}

\def\ul#1{\underline{#1}}
\def\proj{{\cal{P}}}
\def\trans{{\cal{T}}}
\def\unit{{\cal{I}}}

\def\aug{\vert R,L,\{\emptyset\}\rangle }
  \def\ket{\vert \vert  \{ \emptyset \} \rangle}
  \def\ket2{\vert \vert \otimes \{ R \} \rangle}

\def\.#1{\mathaccent 95#1}
\def\^#1{\mathaccent 94 #1}
\def\~#1{\mathaccent "7E #1}

\def\equal{\enskip =\enskip}
\def\plus{\enskip +\enskip}
\def\minus{\enskip -\enskip}
\def\eq{\enskip =\enskip}
\def\pls{\enskip +\enskip}
\def\mns{\enskip -\enskip}

\def\ul#1{\underline{#1}}

\def\c#1{\mbox{\bf #1}}

\def\unit{{\cal I}}
\def\trans{{\cal T}}
\def\proj{{\cal P}}

  \def\proj{{\cal P}}
  \def\trans{{\cal T}}
  \def\ket{\vert \vert  \{ \emptyset \} \rangle}
  \def\ket2{\vert \vert \otimes \{ R \} \rangle}

\documentstyle[aps,epsfig]{revtex}
\begin{document}
\twocolumn[\hsize\textwidth\columnwidth\hsize\csname @twocolumnfalse\endcsname
\title{\bf Magnetic properties of disordered CoCu alloys :  A first principles approach} 
\author{Subhradip Ghosh\cite{ghosh} and Abhijit Mookerjee} 
\address{S.N. Bose National Centre for Basic Sciences, \\
 JD Block,  Sector 3,  Salt Lake City,  Calcutta 700091, 
India.}
\maketitle
\vskip 1.0cm
\begin{abstract}
Crystalline Co$_{x}$Cu$_{1-x}$ alloys show interesting magnetic behavior over the
entire concentration regime. We here present a fully self-consistent first principles
electronic structure studies of the electronic structure and magnetic properties of
the system.We present results for the variation of density of states, magnetic moment, 
spin susceptibility and Curie temperature.
\end{abstract}
\vskip 0.2cm 
\noindent PACS no : 71.20, 71.20c
\vskip 0.2cm
\noindent Keywords :  Magnetism, Alloys, Augmented Space Recursion
\vskip 2.0cm
]

\section{Introduction}
 CoCu alloys have been extensively studied experimentally.
 Earlier investigations, particularly those of the magnetic properties,
  include high-temperature magnetic moment  for 40-
 85\% Cu rich Co-Cu solid solutions \cite{kn:kneller} and variation of
 magnetic moments for Co rich alloys at room temperature \cite{kn:ben,kn:crangle}.
 Recently, the focus of attention has been  on giant 
 magneto-resistance (GMR) studies. Several experimental groups have been carrying 
 studies  of GMR in this system. Most remarkable works
 include investigation of GMR in bulk Co$_{x}$Cu$_{1-x}$ alloys for x = 0.05-0.2
 \cite{kn:yu,kn:xiao},  in heterogeneous thin film
 CoCu alloys \cite{kn:berk,kn:berk1} and in {\sl as}-grown, epitaxial Co-Cu alloy 
layers \cite{kn:park}. Other magnetic studies
 on this system include magnetic anisotropy  of Co on Cu(110) \cite{kn:fass}
 and of Co-Cu ultra thin films \cite{kn:hope}. An altogether different type of work 
 by Childress and Chien \cite{kn:child} is, to our knowledge, the only one of its kind 
  which involved low-temperature magnetic studies of Co$_{x}$Cu$_{1-x}$ alloys.
 This work is worth mentioning because it revealed very interesting features of 
 the magnetic phases over the entire concentration range from x = 0 to x= 0.8. It indicated 
a low-temperature spin glass phase upto x = 0.23 ; a low-temperature 
 re-entrant spin-glass or mixed phase,  stable in the range  0.24 $<$ x $<$ 0.40 
and the normal low-temperature long-ranged random ferromagnetic phase stable beyond 
x = 0.4. 
   Mandal and Ghatak \cite{kn:mand} have reported calculations on 
 the periodic Anderson model for a binary alloy, working within the  
 Hartree-Fock, coupled with the Virtual Crystal Approximation (VCA). They
 have been able to produce some of the features observed experimentally. The model 
used by these authors have many empirical fitted parameters. The existence of a 
spin-glass and, at higher concentrations, a re-entrant mixed phase has been known from
several mean-field approaches \cite{kn:mr,kn:gt}. To our
 knowledge, the system has never been investigated from a first-principles electronic
 structure point of view. The aim of this communication is to present such a calculation of 
 magnetic properties of the fcc, ferromagnetic phase of Co$_{x}$Cu$_{1-x}$
 alloys in the concentration range from 40\% to 80\%
 of Co. We have used the fully self-consistent, tight-binding linearized muffin-tin
orbitals augmented space recursion (TB-LMTO-ASR) technique as the basis of  our calculations.
 We have restricted ourselves to the ferromagnetic phase only leaving the spin-glass
and mixed phases for a later study. Our studies involved calculation
 of magnetic moments, Curie temperatures and the spin susceptibilities.

 \section{Theoretical Details}

\subsection{The magnetic phases}
Description of  magnetic phases within the local spin density approximation
(LSDA) involves the study
of the evolution of local magnetic moments in the vicinity of ion
cores because of the distribution of the valence electron charge.
Each lattice site in the face centered cubic structure is occupied
by an ion core : in our case randomly by either Co or Cu. We shall
 associate a cell or a sphere  with each ion core  and assume that the charge
 contained in the sphere belongs to that ion
core alone. Ideally such cells or spheres should not overlap.
In the traditional Kohn-Korringa-Rostocker (KKR) method this
is certainly so. However,  in the atomic sphere approximation (ASA) 
which we shall use in our TB-LMTO version,  this  
division of space is to a certain extent arbitrary.  Within these
cells the valence electrons carrying spin $\sigma$ sees a binary
random spin-dependent potential V$^{\lambda}_{\sigma}$($\underline{r}$), 
where $\lambda$\equal Co or Cu and $\sigma$\equal $\uparrow$ or $\downarrow$.

The charge density within the cells can be obtained from the
partially averaged Green functions :

\begin{eqnarray}
\~\rho_{\sigma}(r)& = & -(1/\pi) \Im m \sum_{L} \int_{-\infty}^{E_{F}}
[x\ll G^{Co,\sigma}_{LL}(r,r,E)\gg \nonumber\\
& + & (1-x) \ll G^{Cu,\sigma}_{LL}(r,r,E)\gg]dE
\end{eqnarray}

where $\ll G^{Co, \sigma}_{LL}(r, r, E)\gg$ and   $\ll G^{Cu, \sigma}_{LL}(r, r, E)\gg$ are partially
averaged Green functions with the site $r$ occupied  by a 
Co or Cu ion core potential corresponding to spin $\sigma$.

For the random ferromagnetic phase we proceed as follows : we consider all
cells to be identical in that they all carry identical average charge densities.
We shall borrow the notation of Andersen {\it etal} \cite{kn:ajs} to write
functions like $\~f(r_R)$ which are equal to $f(r)$ when $r$ lies
in the atomic sphere labelled by $R$ and is zero outside. The ferromagnetic charge densities
are defined as :

\begin{eqnarray*}
\rho_{1}(r)\equal \sum_{R} \~\rho_{\uparrow}(r_R) 
 \\
\rho_{2}(r)\equal \sum_{R} \~\rho_{\downarrow}(r_R) 
\end{eqnarray*}

The magnetic moment  per cell (atom) is then defined by :

\begin{eqnarray*}
 m & = & (1/N) \int d^{3}\ul{r} \left[ \rho_{1}(r)\minus
\rho_{2}(r)\right] \\
   & = & (1/N) \sum_{R} \int_{r\le S} d^{3}r \left[ \~\rho_{\uparrow}(r_{R})
\minus \~\rho_{\downarrow}(r_{R})\right] \\
& = & (1/N) \sum_{R} \int_{r\le S} d^{3}r\enskip
 m_{R}(r_{R}) \\
\end{eqnarray*}

Since all cells are identical,  the above calculation need be done only
in one typical cell. Within the TB-LMTO-ASA the cells are replaced by
inflated atomic spheres and the remaining interstitial is neglected. The 
problem is then one of a binary alloy with an almost non-magnetic charge density
due to the Cu ion cores and a magnetic one due to the Co ones.
The averaging is done over configurations of the random alloy.

\subsection{The configuration averaging and TBLMTO-ASR}
 For random alloys we extract physical properties using configuration averaging
which means that the physical quantities involved are the configuration averaged
quantities.
 A powerful technique of carrying out this averaging is the augmented space recursion
\cite{kn:asr}. The method allows us to go well beyond the traditional
single site coherent potential approximations and has been applied
successfully to a wide variety of systems \cite{kn:pm,kn:dm,kn:bs,kn:bs1}. The convergence
of the ASR has been established recently \cite{kn:gdm},  so that any
approximation we impose on the recursion is controlled by tolerance
limits preset by us. ASR coupled with TB-LMTO-ASA has been proved to be a very powerful
technique in predicting material properties.
 
 The TB-LMTO-ASR has been described in great detail  earlier \cite{kn:asr,kn:pm,kn:dm,kn:bs,kn:gm}
 . We shall refer the reader to the referenced monograph for technical
 details.Here we quote only the main results.

Our starting point is the TB-LMTO Hamiltonian in atomic sphere approximation (ASA) :

\[ H^{(\gamma)}\eq E_{\nu}\plus hI \] 
\vskip 0.2cm

where, $ h\equal  C\mns E_{\nu}\pls \Delta^{1/2} S \Delta^{1/2}$.
\vskip 0.2cm

$C$,  and $\Delta$ are diagonal matrices in angular momentum space and are th
e potential parameters
of the TB-LMTO technique and $S$  is the structure matrix which is sparse in the
 most tight-binding representation.

 Let us now look at the transformation of the Hamiltonian in the full augmented  space
when the potential parameters have homogeneous binary random distributions.  Such  random
 variation may be described by random site occupation variables $n_{R}$ which take
 values 1 or 0 according to whether the site $R$ is occupied by an A-type
 or B-type
 of atom and have probability densities

 \[ p(n_{R})\eq  x \;\delta(n_{R}-1)\pls(1-x)\;\delta(n_{R})  \]

The details of construction  of the augmented space and the effective Hamiltonian on it has
been  described in detail in our earlier
communications \cite{kn:dm,kn:mook,kn:mook2}. We shall quote
 here only the final result and refer the readers to the referenced works
. The  trick involves
first in replacing each random parameter $P$ in the Hamiltonian by expressions of the type :

\begin{equation} P_{A}\; n_{R}\pls P_{B}\; (1-n_{R}) \eq P_{B}\pls (P_{A
}-P_{B})\; n_{R} \end{equation}

 and then replacing  each factor $n_{R}$ by an operator ${\cal M}_{R}$ which acts on the configuration
space of $n_{R}$.  For a binary distribution of $n_{R}$ the rank of this
space is two. We
shall construct a basis in this space which we shall designate as $\{ \vert\uparrow\rangle\; , \;
\vert\downarrow\rangle\} $. For the binary distribution this operator is

\[ {\cal M}_{R} \equal x\proj_{\uparrow}^{R}\pls (1-x)\proj_{\downarrow}
^{R}\pls \sqrt{x(1-x)}  \left\{
\trans_{\uparrow\downarrow}^{R}\pls\trans_{\downarrow\uparrow}^{R}\right
\} \]

and the above equation becomes  :

\begin{eqnarray*}
& \eq& \c{A}(P)\unit \pls \c{B}(P)\proj_{\downarrow}^{R}\pls\c{F}(P) \left\{\trans_{\uparrow\downarrow}^{R}+\trans_{\downarrow\uparrow}^{R}\right\} \\ & \eq & \^P
\end{eqnarray*}

where

\begin{eqnarray*}
\c{A}(P)& \eq & x\; P_{A}\pls (1-x)\; P_{B} \\
\c{B}(P)& \eq & (1-2x)\;(P_{A}-P_{B})\\
\c{F}(P)& \eq & \sqrt{x(1-x)}\;(P_{A}-P_{B})
\end{eqnarray*}

We  now proceed as  follows  :

\begin{eqnarray}
( E\mns H)^{-1}  & = & ( E\mns C\mns \Delta^{ 1/2}S\Delta^{1/2})^{-1} \nonumber\\
& = & \Delta^{-1/2}\left[ \frac{E-C}{\Delta} - S \right]^{-1}\Delta^{-1/2}
\end{eqnarray}

All factors except $S$ have binary random distributions. Using the procedure described
above we may now convert  the above equation into augmented space. The augmented space theorem
gives \cite{kn:mook}

\[
\ll G_{RL, RL}(E) \gg\eq \langle R, L, \{\emptyset\}\vert
\left(E\^I\mns \^ H\right)^{-1}\vert R, L, \{\emptyset\}\rangle
\]

First  note that :

\begin{eqnarray}
\~\Delta{^{-1/2}}\aug & =& \c{A}(\Delta^{-1/2})\aug \nonumber\\
& +& \c{F}(\Delta^{1/2})\vert R, L, \{R\}\rangle \eq \vert 1\}
\end{eqnarray} 

and if we define $\left[ \c{A}(1/\Delta)\right]^{1/2}\vert 1\}$ as $\vert 1\rangle$,  then this latter
ket is normalized. A little algebra then gives :

\begin{equation}
\ll G_{RL, RL}(E)\gg\eq \langle  1\vert [ E-\^A+\^B+\^F-\^S
]^{-1} \vert 1\rangle
\end{equation}

where

\begin{eqnarray}
\^A &  \eq& \left\{\c{A}(C/\Delta)/\c{A}(1/\Delta)\right\} \unit\otimes\unit\otimes\unit \nonumber\\
\^B & \eq & \left\{ \c{B}((E-C)/\Delta)/\c{A}(1/\Delta)\right\} \sum_{RL
}\proj_{R}\otimes\proj_{L} \otimes\proj^{R}_{\downarrow} \nonumber\\
\^F &\eq & \left\{ \c{F}((E-C)/\Delta)/\c{A}( 1/\Delta)\right\} \sum_{RL
}\proj_{R}\otimes\proj_{L}\nonumber\\ & \otimes &\{\trans_{\uparrow\downarrow}^{R}+\trans_{\downarrow\uparrow}^{R
}\}  \nonumber\\
\^S & \eq & \sum_{RL}\sum_{R'L'}  \left\{ \c{A}(1/\Delta)^{-1/2} \right\}
S_{RL, R'L'} \left\{ \c{A}(1/\Delta)^{1/2}\right\} \trans_{RR'}\nonumber\\& \otimes &\trans_{LL'}\otimes\unit
\end{eqnarray}

and $\^J\eq \^J_{A}\pls \^J_{B}\pls \^J_{F}$ 
where the operators on the right hand side  of these
 equations are given by

\begin{eqnarray*}
\^J_{A} &  \eq& \left\{\c{A}\left((C-E_{\nu})/\Delta\right)/
\c{A}(1/\Delta)\right\} \;\unit\otimes\unit\otimes\unit \nonumber\\
\^J_{B} & \eq & \left\{ \c{B}\left((C-E_{\nu})/\Delta\right)
/\c{A}(1/\Delta)\right\} \;\sum_{RL}\proj_{R}\otimes\proj_{L}
\otimes\proj^{R}_{\downarrow} \nonumber\\
\^J_{F} &\eq & \left\{ \c{F}\left((C-E_{\nu})/\Delta\right)/
\c{A}( 1/\Delta)\right\}
\;\sum_{RL}\proj_{R}\otimes\proj_{L}\nonumber\\
& \otimes &\{\trans_{\uparrow\downarrow}^{R}+\trans_{\downarrow\uparrow}^{R
}\}  \nonumber\\
\end{eqnarray*}

This form is suitable for recursion. We start the recursion with the 
{\sl state} $\vert 1 \rangle$
which is a mixture of the configurations $\{\emptyset\}$ and $\{R\}$ and
carry out recursion as usual.

The initial TB-LMTO potential parameters are obtained from   suitable guess potentials
as described in the article by Andersen {\it etal} \cite{kn:ajs}. In subsequent iterations
the potentials parameters are obtained from the solution of the Kohn-Sham equation

\begin{equation}
\left\{ -\frac{\hbar^{2}}{2m} \nabla^{2} + V^{\nu\sigma} - E\right\} \phi^{\nu}_{\sigma}(r_{R}, E)
\; =\; 0 \end{equation}

where, 

\begin{equation}
V^{\nu\sigma}(r_{R})\;  = \; V_{core}^{\nu\sigma}(r_{R}) + V_{har}^{\nu\sigma}(r_{R})
                           + V_{xc}^{\nu\sigma}(r_{R}) + V_{mad}
\end{equation}

\noindent here $\nu$ refers to the species of atom sitting at $R$ and $\sigma$ the spin component.
The electronic position within the atomic sphere centered at $R$ is given by $r_{R}$ =$r-R$.
The core potentials are obtained from atomic calculations and are available for
most atoms.
The treatment of the Madelung potential in a random alloy has always raised
problems. We adopt a procedure suggested by Drchal {\it etal} \cite{kn:dkw} and 
regularly used in CPA calculations within the TB-LMTO.
 We choose the atomic sphere
radii of the components in such a way that they preserve the total volume on the
average and the individual atomic spheres are almost neutral. This ensures
that total charge is conserved,  but each atomic sphere carries no excess
charge. However,  we are careful that such a choice does not violate the
overlap criterion of Andersen and Jepsen \cite{kn:ajs}.  In the ASR-LSDA self-
consistency loop,  charge transfer takes place between these spheres ; however,  at
the end of the self-consistency iterations,  the spheres are approximately
neutral and hence do not contribute to a Madelung energy. This prescription
is to an extent {\sl ad hoc},  and there is no guarantee in general that we
will be able to find such atomic sphere radii. However,  the procedure has
proved rather successful in many earlier CPA \cite{kn:dkw}  and ASR \cite{kn:dsmd} 
 calculations on magnetic
alloys and we shall adopt it here. 
As in the CPA calculations we iterate until the total energy and moments of the charge density
converge. In this sense our calculations are self-consistent in the LSDA sense. 
\subsection{Curie temperature}
We have estimated the Curie temperature by two different methods-Bragg
Williams approximation and Mohn-Wolfharth model \cite{kn:mw}. Both the
procedures are described below:
\begin {enumerate}
\item{\bf Bragg-Williams approximation:}

The estimates of the Curie temperature were obtained from the magnetic pair energies
\cite{kn:mook2}
The pair energies are defined as follows : 
At two sites labeled $r$ and $r'$ in a completely random paramagnetic
background,  we replace the potential by that
of either the up-spin ferromagnetic Co or the down-spin one. The Green function
of this system we shall denote by : $G_{LL}^{Co, \sigma\sigma '}(r, r, E)$,  
$\sigma$ being the spin type
at the site $r$ (either $\uparrow$ or $\downarrow$) and $\sigma '$ that at the site $r'$. The
pair energy is defined as 

\begin{eqnarray}
 E(R) &\eq &\int_{-\infty}^{E_{F}} dE\;E\;[-(1/\pi)\Im m (
G_{LL}^{Co, \uparrow\uparrow}(r, r, E)\nonumber\\ & + &G_{LL}^{Co, \downarrow\downarrow}
(r, r, E) -G_{LL}^{Co, \uparrow\downarrow}(r, r, E)\nonumber\\ & -& G_{LL}^{Co, \downarrow\uparrow}(
r, r, E) )]
\end{eqnarray} 

Here $R\eq r-r'$.
We may either estimate the above directly,  or to be more accurate we may use the
orbital peeling method of Burke \cite{kn:bk}. The latter is an extension of the
recursion method,  where small differences of large energies (as in the definition
of the pair energy) are obtained directly and accurately from the recursion continued
fraction coefficients. Note that we have assumed that the dominant contribution
to the pair energy comes from the band contribution and the rest approximately cancel out. The
simplest Bragg-Williams estimate of the Curie temperature is 

\[ T_{c} \eq (1-x)E(0)/\kappa_{B} \]  where 

$E(0) = E(q=0)$ and 
$E(q)=\sum_{R} \exp{\left(iq.R\right)}E(R)$. Since the pair energy is short-ranged,  a
reasonable estimate of $E(0)$ is $ \sum_{n<3} Z_{n}\;E(R_{n})$ where $R_{n}$ is the n$^{th}$-nearest

neighbour vectors and $Z_{n}$ is the number of n$^{th}$-nearest neighbours. The Bragg-Williams
approach overestimates the Curie temperature and its generalization,  the
cluster variation method,  yields better quantitative estimates. We have restricted
ourselves to the Bragg-Williams nearest neighbour pair energy approximation.

\item{\bf Mohn-Wohlfarth model(MW):}

In this model \cite{kn:mw}the Curie temperature is given by:

$$  \frac{T_{C}^2}{T_{C}^{S^2}}+\frac{T_{C}}{T_{SF}} \mns 1 \eq 0$$

where, 

T$_{c}^S$ is the stoner Curie temperature given by, 

$I(E_{F})\int_{-\infty}^{\infty} N(E) (\frac{\delta f}{\delta E} ) dE \eq 1$

I(E$_{F}$) is the Stoner parameter obtained from the earlier calculations \cite{kn:jan}
, N(E) is the density of states per atom per spin \cite{kn:gun}
and f is the Fermi distribution function.

T$_{SF}$ is the spin fluctuation temperature given by, 

$$ T_{SF} \eq \frac{m^2}{10k_{B}\chi_{0}} $$

$\chi_{0}$ is the exchange enhanced spin susceptibility at equilibrium and m is the magnetic
moment per atom. 

$\chi_{0}$ is calculated using the relation by Mohn \cite{kn:mw} and Gersdorf \cite{kn:ger}:

\[\chi_{0}^{-1} \eq  \frac{1}{2\mu_{B}^2}\left(\frac{1}{2N^\uparrow(E_{F})}\pls 
\frac{1}{2N^\downarrow(E_{F})} \mns I\right)\]

I is the stoner parameter and N$^\uparrow(E_{F})$ and N$^\downarrow(E_{F})$ are the spin-up and spin-down density of states per atom.
\end{enumerate} 
\section{Computational Details}
For the calculation of the component projected averaged density of states
of the ferromagnetic
 phase we have used a real space cluster of 400 atoms and
an augmented space shell upto the sixth nearest neighbour from the starting
state. Eight pairs of recursion coefficients were determined exactly and the
continued fraction terminated by the analytic terminator due to Luchini
and Nex \cite{kn:ln}. In a paper Ghosh {\it etal} \cite{kn:gdm} have shown the
convergence of the related integrated quantities,   like the Fermi energy, 
the band energy,  the magnetic moments and the charge densities,  within the augmented space
recursion. The convergence tests suggested by the authors were carried
out to prescribed accuracies. We noted that at least eight pairs
of recursion coefficients were necessary to provide Fermi energies
and magnetic moments to required accuracies. We have reduced the computational burden of the
recursion in the full augmented space by using the local symmetries of the
augmented space to reduce the effective rank of the invariant subspace
in which the recursion is confined \cite{kn:asr} and using the seed
recursion methodology \cite{kn:gm} with fifteen energy seed points
uniformly across the spectrum.

We have chosen the Wigner-Seitz radii of the two constituent atoms Co
and Cu in such a way that the average volume occupied by the atoms is
conserved. Within this constraint we have varied the radii so that
the final configuration has neutral spheres. This eliminates the
necessity to include the averaged Madelung Energy part in the total
energy of the alloy. The definition and computation of the Madelung
Energy in a random alloy had faced controversy in recent literature
 and to this date no satisfactory resolution of the problem
exists. Simultaneously we have made sure that the sphere overlap remains
within the 15$\%$ limit prescribed by Andersen.

The calculations have been made self-consistent in the LSDA sense,  that
is,  at each stage the averaged charge densities are calculated from the
augmented space recursion and the new potential is generated by the
usual LSDA techniques. This self-consistency cycle was converged in
both total energy and charge to errors of the order 10$^{-5}$. We have
also minimized the total energy with respect to the lattice constant.
The quoted results are those for the minimum configuration. No short
ranged order due to chemical clustering has been taken into account
in these calculations,  nor any lattice distortions due to the size
differences between the two constituents.
\section{Results and Discussion}
\par Figure 1(a) and 1(b) show the partial density of states at Co and
at Cu sites respectively for various concentrations of Cu. It is evident
from the figures that there is hardly any difference in the relative heights
and shapes in the majority and minority spin bands at Cu site for various
concentrations. This is reflected in the local magnetic moment shown in Figure
2 where Cu sites have a very small amount of induced moment. For the minority
spin partial densities on Co, the structure below the Fermi level do not change
much with the Co concentration. The peak at around 0.0 Ryd. grows with decreasing
Co concentration. However, this structure is above Fermi level and does not
contribute to the magnetic moment. For the majority spin partial densities on Co, 
the peak around -0.4 Ryd. doesn't change with concentration while the one around
-0.3 Ryd. grows with concentration. The most remarkable change occurs in the peak
around -0.2 Ryd. whose height reduces with the concentration and finally the peak
around -0.3 Ryd. contributes more with increasing concentration.
As a result,  the partial magnetic moment on Co increases only slowly with concentration
and the variation of average magnetic moment is linear as shown in Figure 2. 

\par Figure 3 shows the variation of average magnetic moment in emu/gm unit and the agreement with
experimental results \cite{kn:child} is pretty good qualitatively. Both CPA and ASR
moments agree well too. But, the agreement with the results from model calculation
\cite{kn:mand} is far from being reasonable.Their calculations predict a monotonic
fall of average magnetic moment and finally vanishing of magnetic moment around 40\%
of Co concentration. This has neither been observed in experiments \cite{kn:child} nor
in our calculations with two widely used {\sl ab-initio} methods. This discrepancy is
due to the choice of technique in case of model calculations. They used Hartree-Fock and
VCA for their calculations. Both the approximations have their own limitations in
predicting properties of real materials. The very important electron-electron correlation
effect is absent within the framework of Hartree-Fock approximation while in LSDA this
effect has been incorporated. The limitation of VCA is that it has been proved to be
successful only in cases of weak scattering alloys where each particle sees nearly the
same average perturbation field. ASR on the other hand is a very powerful and accurate
technique in the sense that it is an exact method involving no mean-field like approximation
for configuration averaging. This is the reason for better agreement of our
calculations with the experimental results.

\par Figure 4 shows the partial and average spin susceptibilities of the system. The partial
susceptibility of Cu varies smoothly and finally becomes almost saturated while we find
little oscillations in case of Co particularly in the range of 60\%-80\% of Co. This
nature is reflected in the average susceptibility. This is basically due to oscillations
in the up density of states at Fermi energy at Co sites(See Table I). 
However, the oscillation in the Co as well as average susceptibilities are very small, 
eventually leading to saturation.

\par Figure 5 shows the variation of Curie temperature with Co concentration 
using Bragg Williams (BW) approximation and MW model. In BW approximation, 
the variation of Curie temperature is smooth though the values are overestimated. 
This is expected as BW approximation has a tendency to overestimate \cite{kn:subh}.
 This overestimation is reduced using MW model but the qualitative agreement between 
the two methods is not obtained. In the concentration regime 40\%-60\% of Co, 
Curie temperature oscillates in MW model and beyond it qualitative
agreement with BW calculations is achieved. This fluctuating behaviour obtained in MW
calculations is due to the oscillations in Spin fluctuation temperature(See Table II)
because Stoner Curie temperature has a systematic variation with Co concentration. This
oscillation in spin fluctuation temperature can be explained as follows: for the concentration 40\%-60\% of Co the partial spin susceptibilities of both Co and Cu undergo variations
which is reflected in the spin fluctuation temperature while for 60\%-80\% of Co the
susceptibilities of both the components remain almost constant except for a small variation
at around 80\% of Co. That's why the spin fluctuation temperature and hence the Curie
temperature follow the same trend as obtained in BW calculations for Co concentration beyond
60\%. The Stoner Curie temperature(See Table II)is very high as expected because LSDA
overestimates the binding.
We have also tried to characterize the system for the concentration regime of our 
investigation following the prescription by Mohn and Wolfharth \cite{kn:mw} which
characterizes a system from the values of Curie temperature and spin fluctuation 
temperature. According to the prescription the quantity t$_{c}$ \eq $\frac{T_{c}}{T_{c}^S}$
determines the agency playing the dominant role in finding out Curie temperature and
hence the effect of spin fluctuations in determining the magnetic properties of the
system. The systems with t$_{c}<$0.5 are called {\sl fluctuation} systems in which 
description of magnetism should involve spin fluctuation effects and systems with t$_{c}>$0.5
are called {\sl Stoner} systems in which single particle excitations play the dominant
role. In our case, we see (See Table II) that for the entire range of concentrations
t$_{c}$ remains less than 0.5 which means that the description of magnetic properties
should have the effect of spin fluctuations.
\section{Conclusions}
We have calculated for the first time, the electronic structure and
magnetic properties of Co$_{x}$Cu$_{1-x}$
alloys using fully self-consistent first principles electronic structure technique 
and obtained results which agree reasonably well with the experimental results. This
clearly shows that ASR coupled with TBLMTO is a powerful technique in describing 
electronic structure and magnetic properties of binary alloys. Our results clearly
show the flaws in the theoretical results based upon model calculations. Finally, this
study clearly presents the fact that in CoCu alloy spin fluctuation effect is very much
present in the ferromagnetic phase and a more accurate description of the magnetic
properties of the system should involve this effect.
\section*{Acknowledgments}
AM and SG acknowledges useful discussions with Dr.Kalyan Mandal. 

\newpage
\section*{Figure Captions}
{\bf Figure 1(a)} shows the partial densities of states of Co at various Co concentrations.
The concentrations are indicated in the inset. Dashed curves show the results for
the minority and full curve the majority spin states. Fermi energies are indicated by
vertical dashed lines with each figure.

\par {\bf Figure 1(b)} shows the partial densities of states of Cu at various Co concentrations.
The concentrations are indicated in the inset. Dashed curves show the results for
the minority and full curve the majority spin states. Fermi energies are indicated by
vertical dashed lines with each figure.

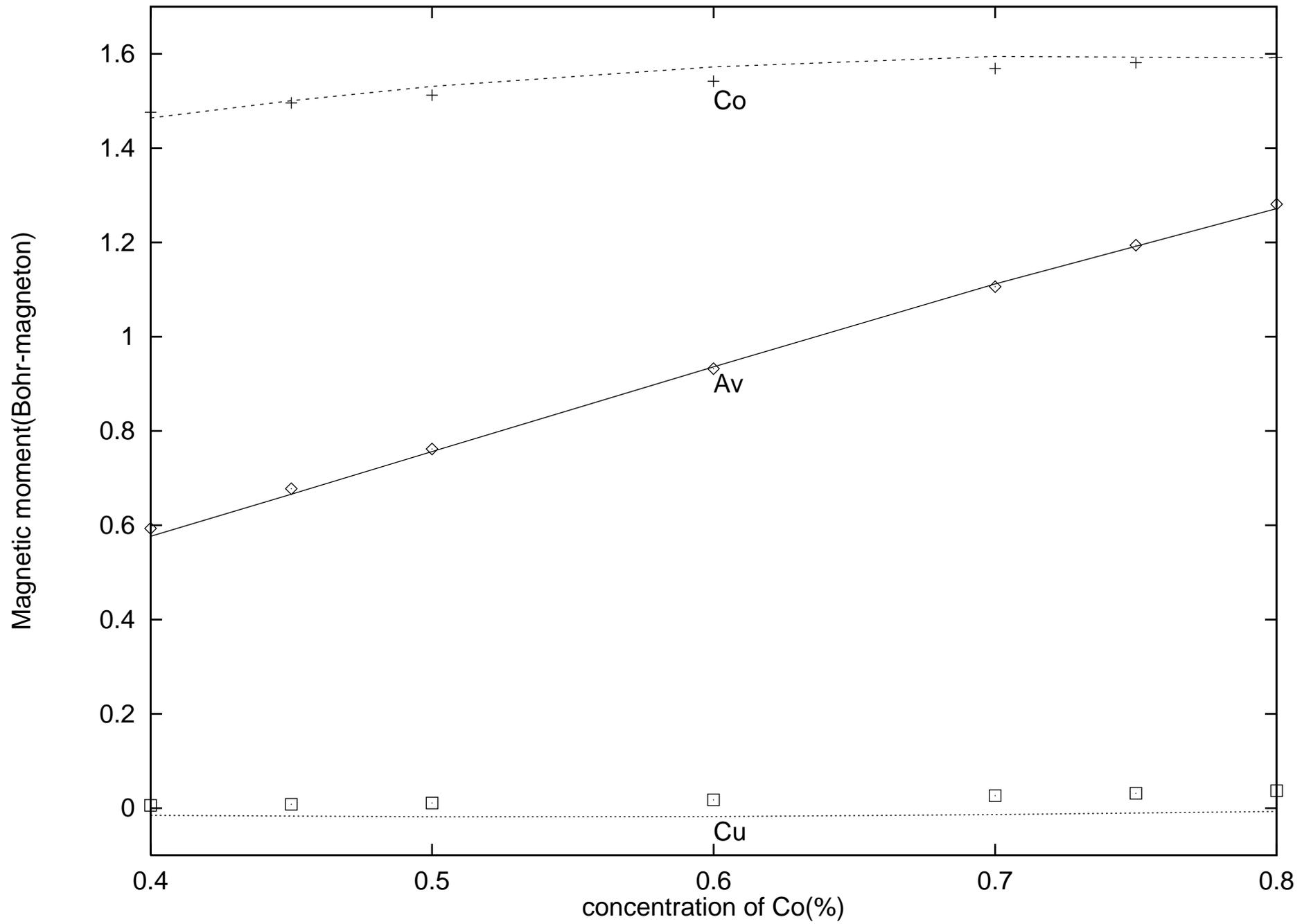
\par {\bf Figure 2} shows The partial magnetic moments on Co and Cu sites
and the average magnetic moment as a function
of Co concentration. The solid and dotted lines refer to the augmented space calculations
while the squares, the crosses and diamonds to the CPA.All these values are in units
of Bohr-magneton.
 
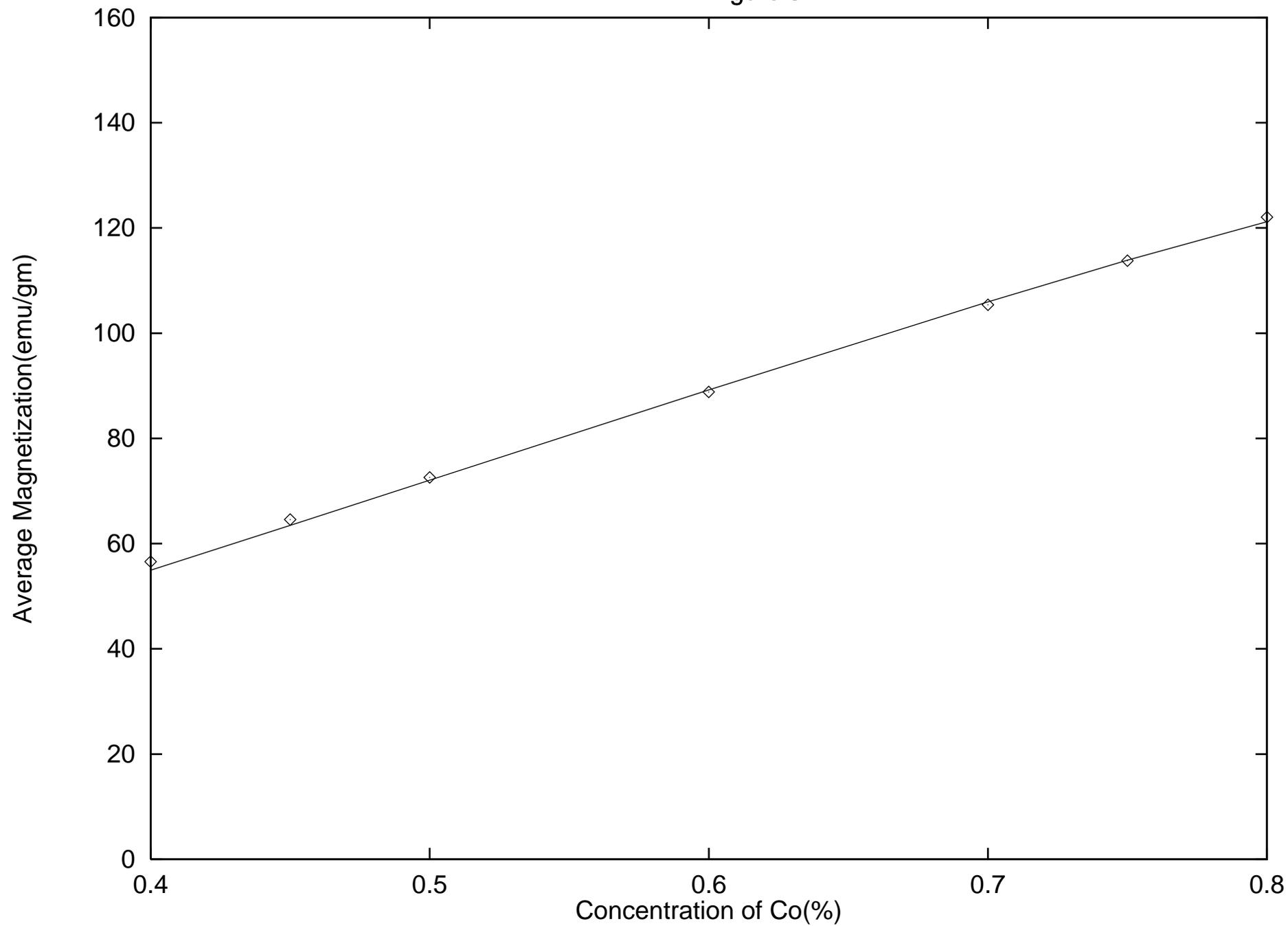
\par {\bf Figure 3} shows the average magnetization in emu/gm as a function of Co concentration
at 0K obtained from the theoretical estimates. The solid line refers to the ASR
calculations while the diamonds to CPA.

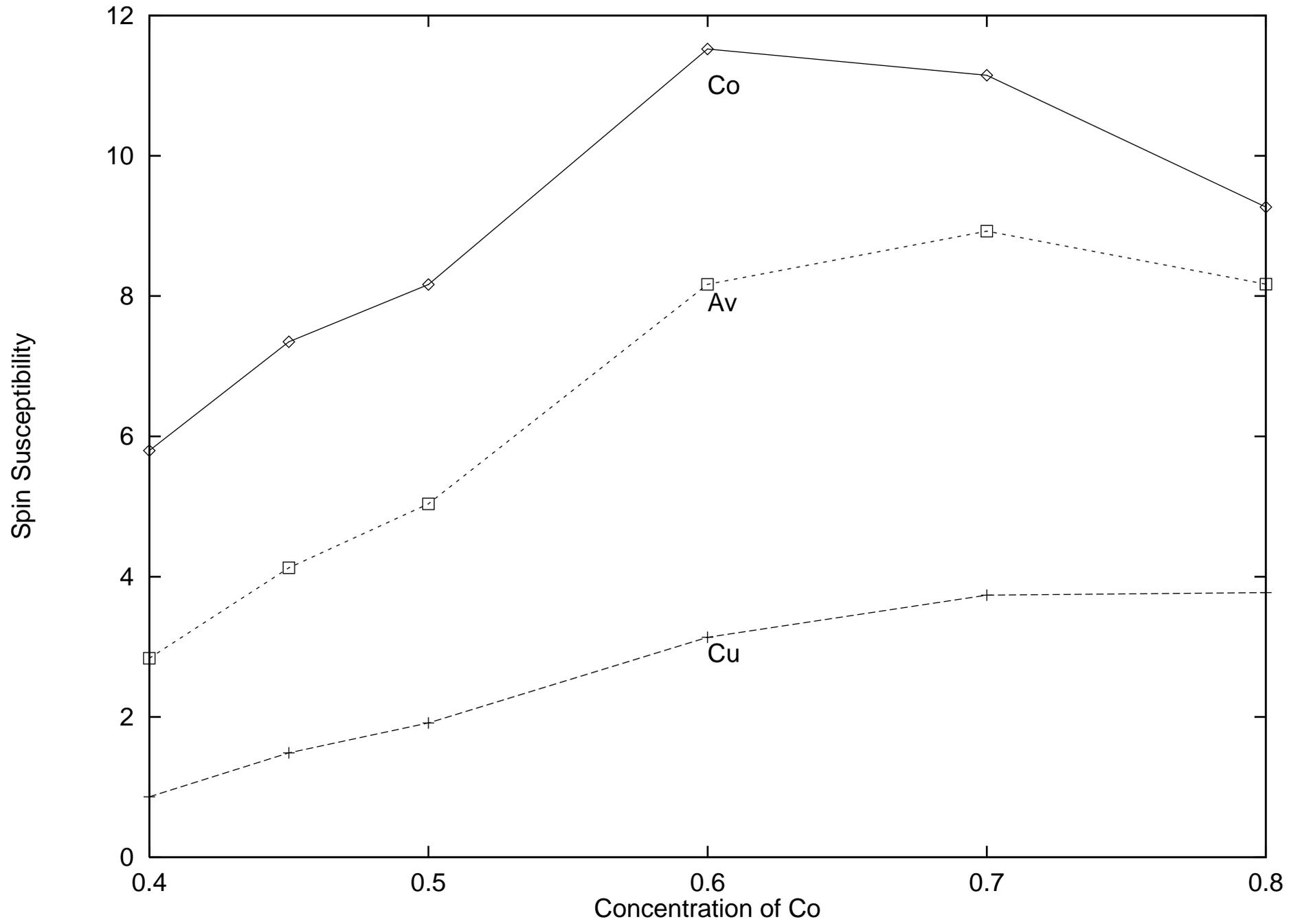
\par {\bf Figure 4} shows the average and partial spin susceptibilities($\frac {\chi}{2\mu_{B}^2}$) 
in Ryd$^{-1}$atom$^{-1}$ with Co concentration.

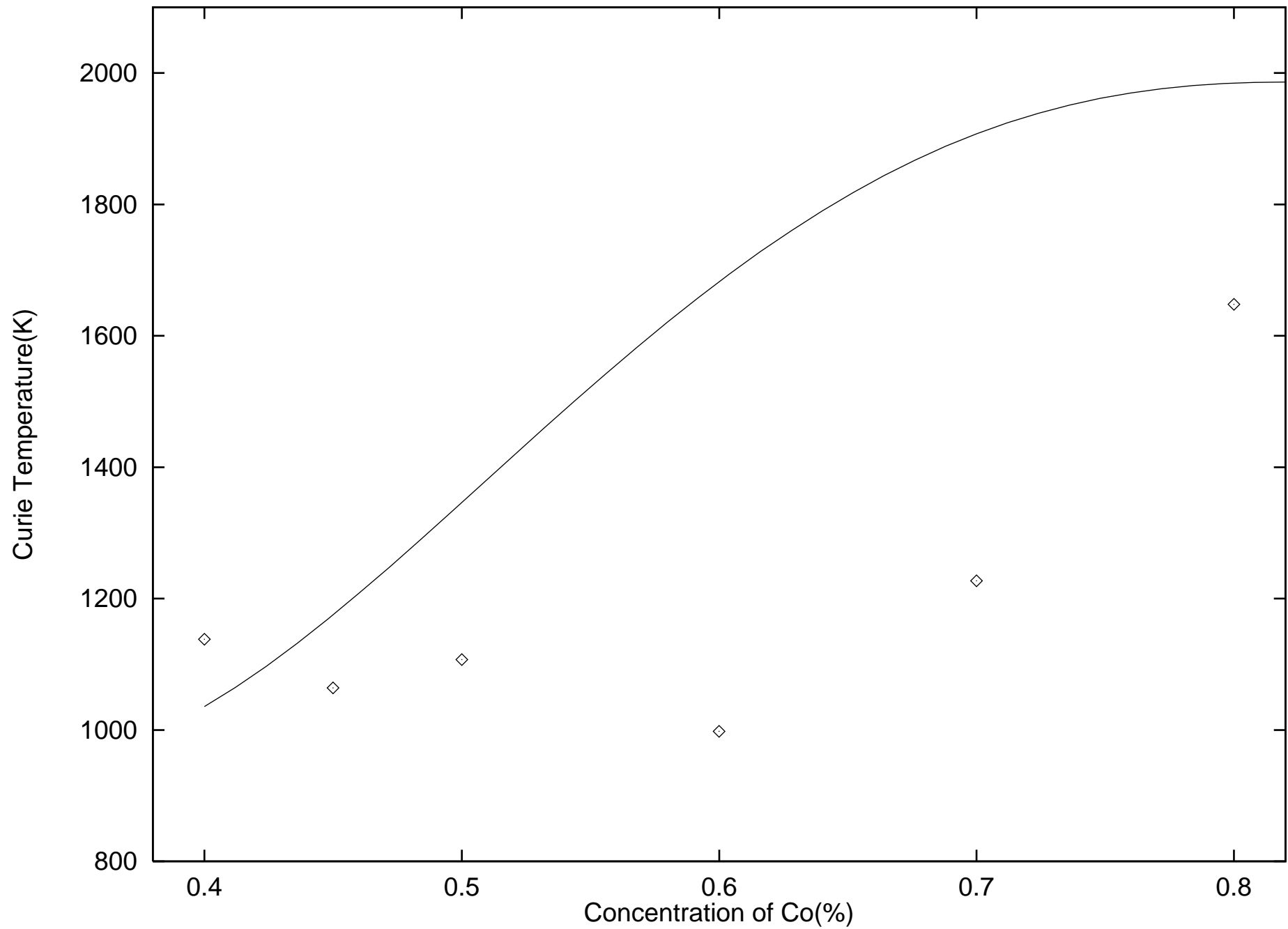
\par {\bf Figure 5} shows the Curie temperature as a function of Co concentration. The solid line
corresponds to the results using BW approximation while the squares correspond to the
results with MW model.
\newpage
\section{Tables}
\begin{table}
\centering
\begin{tabular} {cccccccc}
\hline
x$_{co}$ &N$_{co}^\uparrow$(E$_{F}$) &N$_{co}^\downarrow$(E$_{F}$) &N$_{cu}^\uparrow$(E$_{F}$) &N$_{cu}^\downarrow$(E$_{F}$) &$\frac{\chi_{co}}{2\mu_{B}^2}$ &$\frac{\chi_{cu}}{2\mu_{B}^2}$ &$\frac{\chi_{av}}{2\mu_{B}^2}$\\
\hline
0.4 &2.13 &51.30 &0.77 &0.88 &5.79 &0.86 &2.84\\
0.45 &2.52 &50.54 &1.30 &1.47 &7.35 &1.49 &4.13\\
0.5 &2.72 &47.73 &1.61 &1.88 &8.16 &1.91 &5.04\\
0.6 &3.43 &38.28 &2.41 &3.07 &11.52 &3.13 &8.17\\
0.7 &3.58 &22.55 &2.50 &4.09 &11.15 &3.74 &8.92\\
0.8 &3.37 &15.83 &2.32 &4.84 &9.27 &3.77 &8.17\\
\hline\end{tabular}
\caption{ shows the density of states (both up and down) at fermi
level for both the components and the partial and average susceptibilities with
Co concentration}
\end{table}
\begin{table}
\centering
\begin{tabular} {cccccc}
\hline
x$_{co}$ &T$_{c}^{BW}$  &T$_{SF}$ &T$_{c}^{S}$ &T$_{c}^{MW}$  &t$_{c}$\\
\hline
0.4      &1042  &1169 &7007 &1138 &0.16\\
0.45     &1150  &1089 &7095 &1064 &0.15\\
0.5      &1394  &1134 &7186 &1107 &0.16\\
0.6      &1667  &1016 &7356 &998  &0.14\\
0.7      &1911  &1260 &7530 &1227 &0.17\\
0.8      &1983  &1728 &7709 &1648 &0.22\\
\hline\end{tabular}
\caption{shows the values of BW Curie temperature(T$_{c}^{BW}$), 
MW Curie temperature(T$_{c}^{MW}$),spin fluctuation temperatures(T$_{SF}$) and
Stoner Curie temperature(T$_{c}^S$) with Co concentration}
\end{table}
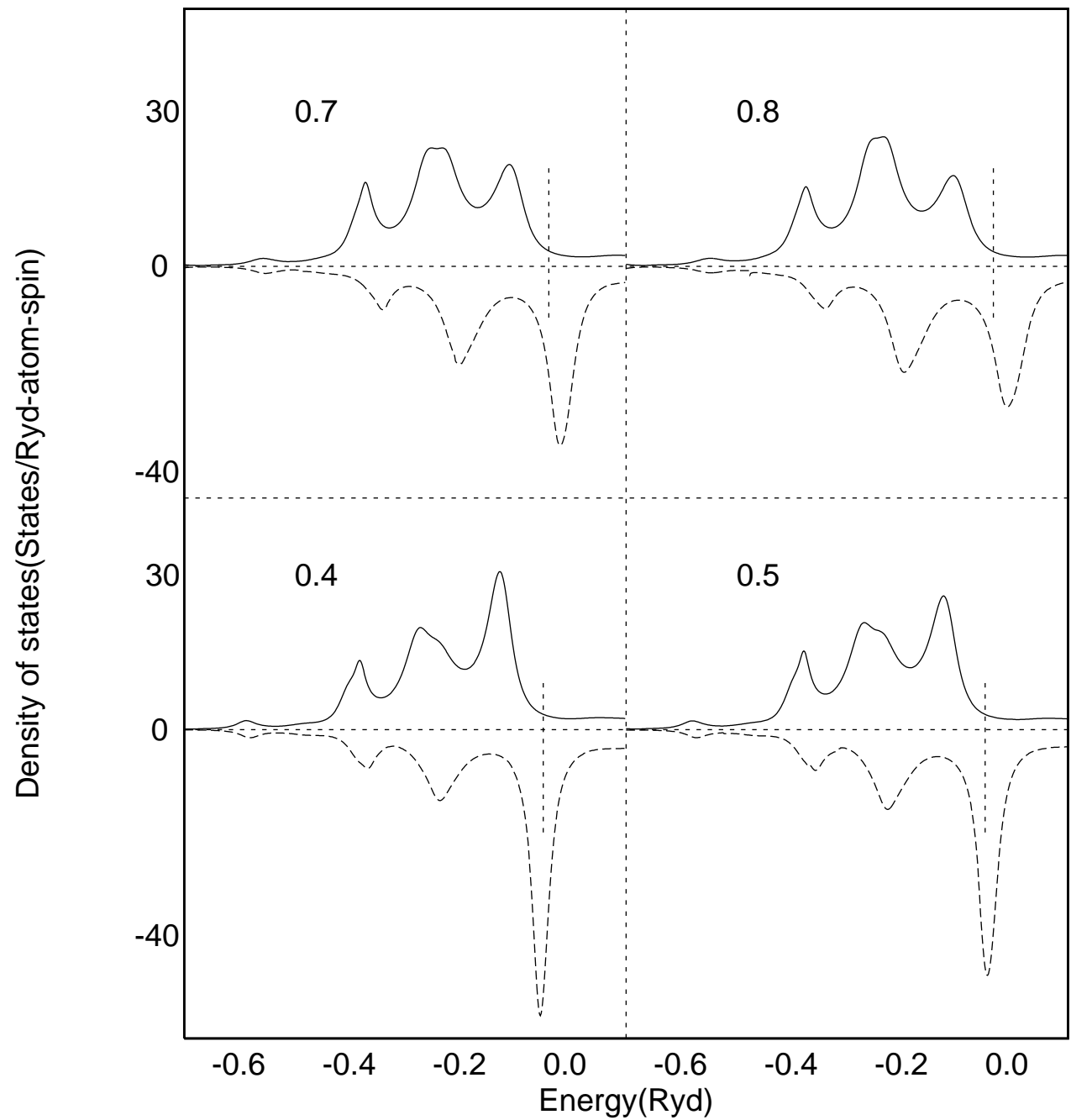
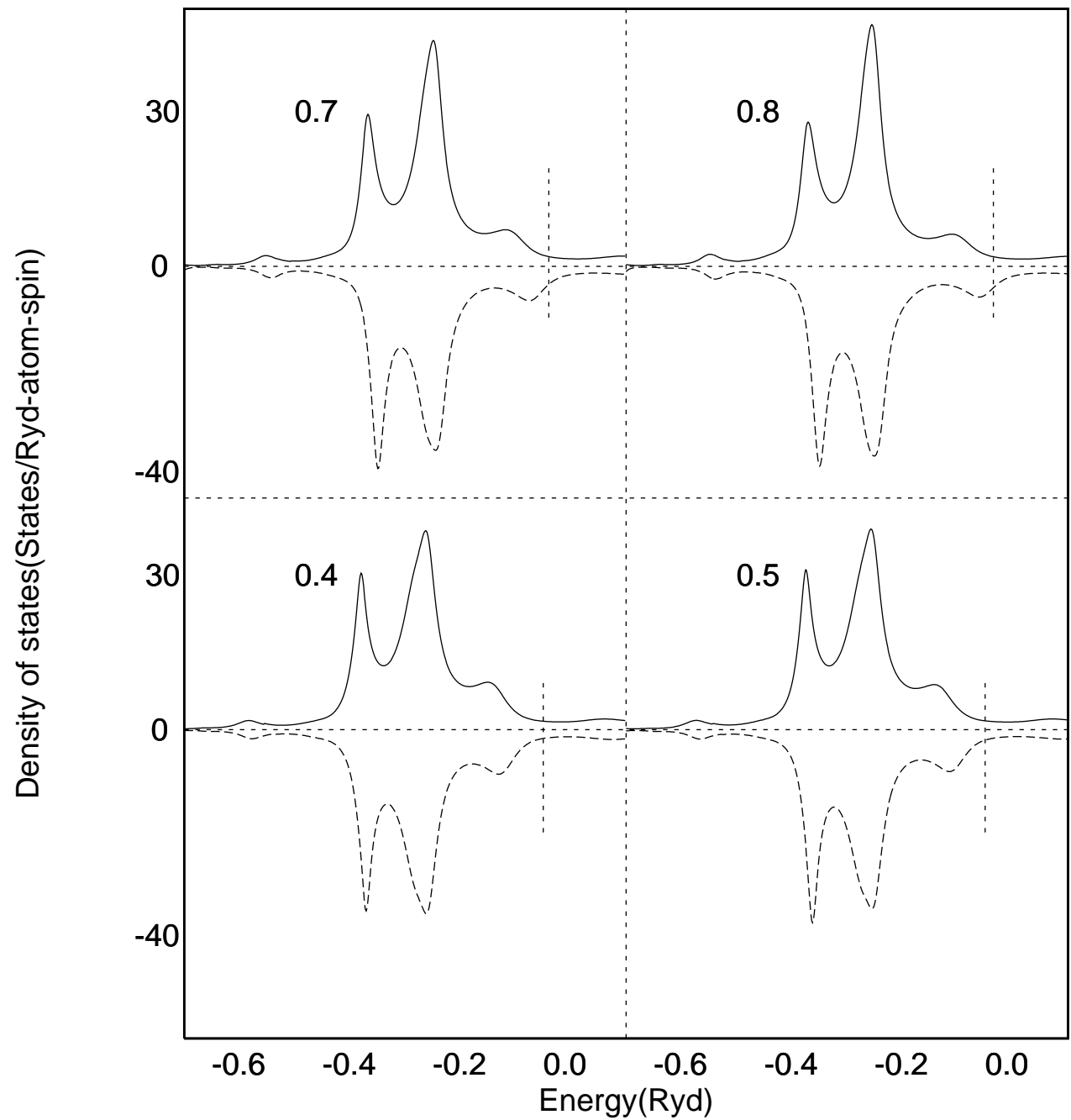
\end{document}